\documentclass[manuscript, screen, authorversion]{acmart}
\AtBeginDocument{%
  }

\setcopyright{acmlicensed}
\copyrightyear{2025}
\acmYear{2025}
\acmDOI{XXXXXXX.XXXXXXX}

\usepackage[
type={CC},
modifier={by-nc-nd},
version={4.0},
]{doclicense}
\begin{document}

\title[Driving Accessibility]{Driving Accessibility: Shifting the Narrative \& Design of Automated Vehicle Systems for Persons With Disabilities Through a Collaborative Scoring System}

\author{Savvy Barnes}
\email{barne329@msu.edu}
\orcid{0009-0003-1167-3737}
\affiliation{%
  \institution{Department of Statistics \& Probability, Michigan State University}
  \city{East Lansing}
  \state{Michigan}
  \country{USA}
}

\author{Maricarmen Davis}
\orcid{0009-0007-6370-3773}
\affiliation{%
  \institution{Department of Electrical Engineering, Michigan State University}
  \city{East Lansing}
  \state{Michigan}
  \country{USA}
}
\email{davis496@msu.edu}

\author{Josh Siegel}
\orcid{0000-0002-5540-7401}
\affiliation{%
  \institution{Department of Computer Science \& Engineering, Michigan State University}
  \city{East Lansing}
  \state{Michigan}
  \country{USA}
}
\email{jsiegel@msu.edu}


\begin{abstract}
  Automated vehicles present unique opportunities and challenges, with progress and adoption limited, in part, by policy and regulatory barriers. Underrepresented groups, including individuals with mobility impairments, sensory disabilities, and cognitive conditions, who may benefit most from automation, are often overlooked in crucial discussions on system design, implementation, and usability. Despite the high potential benefits of automated vehicles, the needs of Persons with Disabilities are frequently an afterthought, considered only in terms of secondary accommodations rather than foundational design elements. We aim to shift automated vehicle research and discourse away from this reactive model and toward a proactive and inclusive approach. We first present an overview of the current state of automated vehicle systems. Regarding their adoption, we examine social and technical barriers and advantages for Persons with Disabilities. We analyze existing regulations and policies concerning automated vehicles and Persons with Disabilities, identifying gaps that hinder accessibility. To address these deficiencies, we introduce a scoring rubric intended for use by manufacturers and vehicle designers. The rubric fosters direct collaboration throughout the design process, moving beyond an ``afterthought'' approach and towards intentional, inclusive innovation. This work was created by authors with varying degrees of personal experience within the realm of disability.
\end{abstract}



\received{20 February 2025}
\received[revised]{12 March 2009}
\received[accepted]{5 June 2009}

\maketitle

\doclicenseThis%

\section{Introduction}
\label{sec:introduction}

Automated vehicles (AVs) and related technologies are evolving rapidly, with daily advancements driving their potential to reshape mobility. For Persons with Disabilities (PwDs), AVs can provide unprecedented levels of independence and access to transportation\cite{claypool_self-driving_2017}. However, AV design and deployment processes often appear to neglect the specific needs of PwDs, perpetuating biases and accessibility gaps within this already marginalized community\cite{darcy_road_2018, hamraie_crip_2019}. Three key factors contribute to this systemic oversight: (1) a fragmented regulatory framework that fails to mandate accessibility as a core requirement, (2) minimal direct collaboration between AV manufacturers and PwDs, leading to solutions that fail to adequately address user needs, and (3) limited research that holistically considers both barriers and benefits of AV technology for PwDs.

Existing literature on AV accessibility either highlights barriers or emphasizes advantages, rarely bridging both perspectives in a meaningful way. Without acknowledging barriers, the significance of advantages is diminished, and without showcasing advantages, barriers lack urgency. This imbalance contributes to public misconceptions about AVs and their regulatory landscape, which in turn affects policies that fail to integrate accessibility concerns.

To address these problems, we propose the Vehicle Accessibility Rating System (VARS), a scoring rubric designed to evaluate AV inclusivity based on direct user input from Persons with Disabilities. Existing accessibility assessments primarily focus on regulatory compliance, but VARS is built on continuous user engagement to ensure meaningful improvements throughout the AV design and deployment process. In this way, we reduce the disconnect between AV systems, PwDs, and the general public to enable the fullest potential of AV systems for those most in need.

The organization of this article will follow a similar format to the 2015 National Council on Disability report titled ``Self-Driving Cars: Mapping Access to a Technology Revolution'' \cite{noauthor_national_nodate}. First, we will discuss the present state of automated vehicles. This will allow us to discuss the potential barriers and advantages of these systems in Section~\ref{sec:baradv}. Section~\ref{sec:regs} covers the current regulations and policies on AVs and PwDs, jointly. This section is divided into global regulations and those specific to the United States of America. Here, we show that current regulations do not meet the needs of PwDs. In response, we propose a new regulatory scoring rubric in Section~\ref{sec:rub} that may foster future communication and accountability in regulations and policies regarding automated vehicles and Persons with Disabilities.

\section{Current State of Automated Vehicle System Technology}
Automated Vehicles, like self-driving cars, are equipped with technology to fully or partially replace a human driver to navigate from an origin to a destination while responding to environmental conditions and avoiding hazards. Interest in self-driving cars has increased in recent years due to their potential to revolutionize transportation worldwide.\cite{HOPKINS2021101488}

Modern vehicles are often equipped with advanced driver assistance systems (ADAS), such as adaptive cruise control, automatic emergency braking, lane-keeping assist, and self-park \cite{web:State_of_ADAS}. The new features and systems have been shown to improve driver and passenger experience\cite{VO_Experiences}; however, disability status is not commonly included as a demographic variable, leading to potential research oversights. 

\subsection{Levels of Automation}
The automotive industry has adopted the Society of Automotive Engineers (SAE) standard J3016 \cite{web:SAE_J3016}, which foramlizes a taxonomy for vehicle driving automation, as shown in Table~\ref{table:Levels_of_Aut}. This taxonomy defines six levels based on the operating capabilities and the potential requirement for human driver intervention, ranging from Level 0 (vehicles without driving automation) to Level 5 (vehicles with full driving automation in all circumstances, never requiring human input); however, this standard does not include mention usability, accessibility, or inclusivity for PwD.

\begin{table}
  \caption{Levels of Driving Automation}
  \label{table:Levels_of_Aut}
  \begin{tabular}{c l p{10cm}}
    \toprule
    Level & Driving Automation & Description \\
    \midrule
    0 & No Automation & The driver must operate the vehicle and remain fully engaged at all times. The vehicle's system produces warnings and alerts but does not control any of the driving aspects. \\
    1 & Driver Assistance & The driver must operate the vehicle and remain fully engaged at all times. The vehicle's system can support some aspects of driving. \\
    2 & Partial Automation & The vehicle's system provides continuous assistance and is fully automated under specific scenarios. The driver must remain fully engaged and ready to take control as needed. \\
    3 & Conditional Automation & The vehicle's system performs all aspects of driving. It is fully automated under most scenarios, but the driver must be available and ready to take control as needed. \\
    4 & High Automation & The vehicle's system performs all aspects of driving. It is fully automated within its operating domain. The occupants act as passengers; no driver is needed to operate the vehicle. \\
    5 & Full Automation & The vehicle is fully automated and does not need a driver to operate the vehicle. The occupants act as passengers. The vehicle's system can operate under all conditions and scenarios. \\
    \bottomrule
  \end{tabular}
\end{table}

No AVs level three or higher are commercially available in the US as of 2024\cite{web:Automated_Vehicles_for_Safety, nhtsa_automated_vehicles}. 

\subsection{Assistive Technology}
PwD often rely on adaptive equipment or after-purchase modifications to ensure their vehicles meet their specific needs. The burden of identifying the right adaptive equipment solutions, consulting the required specialist, and completing the evaluations needed is placed on PwD and/or their caregivers \cite{web:Adapted_Vehicles}.

Some initiatives focus on making AVs more accessible to PwD, such as the Purdue University EASI RIDER (Efficient, Accessible, and Safe Interaction in a Real Integrated Design Environment for Riders with Disabilities). The EASI RIDER project \cite{web:Purdue_EASIRIDER} aims to integrate accessibility features (wheelchair ramps, voice-activated features, control interfaces via mobile devices, etc.) tailored to the needs of PwD. 

Some vehicle manufacturers have made attempts to create accessible vehicles; however, increasing costs and limited demand led most to abandon efforts\cite{National_Museum}. Where attempts are still being made, as such with Ford Motor Company , accessibility is based on adaptations to pre-existing vehicles \cite{Ford_Motor_Company}. Although programs such as these are still beneficial, they do not solve all the problems that accessible design and communication can solve. Additionally, adaptive technology is not widely available like autonomous vehicle systems are becoming; therefore, it has diminished Curb-Cut Effect, where accessible design ends up helping all people and not only those who need them \cite{Blackwell_2017}.

While progress is being made, continuous dialogue between manufacturers, policymakers, disability advocate groups, and PwD is essential to ensure that automated vehicles can fully serve the diverse needs of disabled people. The Disability Rights Education \& Defense Fund \cite{web:DREDF} emphasizes that AVs must be designed with accessibility as a core feature, not an afterthought, a perspective shared by the authors.

\section{Advantages \& Barriers}
\label{sec:baradv}
Contemporary literature tends to focus on the barriers or the advantages of automated vehicles and Persons with Disabilities, but rarely are they considered jointly. If they are, the emphasis is commonly placed on barriers; e.g., in Cordts et al. \cite{CORDTS2021101131}, mobility barriers and perceptions of AVs are discussed for persons with physical disabilities, with only brief mention of how AV systems can address these barriers. Similarly, Harper et al. \cite{harper_estimating_2016} mention the critical need for AV systems to provide PwDs access to healthcare, but takes a deeper dive into the barriers for the general public posed by an increase in automated transportation.  

A discussion of advantages is crucial to motivate efforts to overcome barriers, just as recognizing barriers prevents advantages from being treated as trivial. When advantages are presented without acknowledging barriers, they often appear as afterthoughts—developed primarily for able-bodied users, with benefits for PwDs emerging incidentally. This ``benefit-afterthought'' perspective discourages proactive accessibility efforts, even though principles like the Curb-Cut Effect \cite{Blackwell_2017} demonstrate that accessible innovations benefit society as a whole.

AV technologies are rarely developed in collaboration with PwDs. Instead, accessibility is often framed as a feature rather than an integral design consideration, with little input from the communities these systems claim to serve. Discussing barriers and advantages together helps shift this narrative, ensuring that accessibility solutions directly address existing challenges rather than being secondary outcomes of general technological progress. Some recent literature has started adopting this approach \cite{darcy_road_2018}.

Recent studies categorize AV-related advantages and barriers as either direct or indirect, which align closely with technological and societal impacts, respectively. This distinction is critical because PwDs face unique challenges and benefits compared to the general public. However, confirmation bias leads many to prioritize information that aligns with their existing knowledge \textendash often excluding disability-related concerns \cite{aba_commission_on_disability_rights_aba_nodate, harvey_rights_2017, lee_hardest_2020}. As a result, PwDs are frequently left out of discussions about AV development, despite the technology’s potential to improve their health, social mobility, and economic participation\cite{claypool_self-driving_2017, darcy_road_2018, hamraie_crip_2019,monteleone_forgotten_2020}. To create an equitable AV landscape, it is essential to consciously address both direct and indirect advantages and barriers, ensuring that technological progress does not overshadow critical accessibility concerns.

Understanding the distinction between direct and indirect factors is essential, as each comes with unique societal, financial, and physical constraints. Direct advantages/barriers are those that affect the ability of a PwD to drive or utilize automated vehicles, whether as a pedestrian, passenger, or driver. They involve improvements or declines within the vehicle system that alter accessibility or ease of use. These typically affect the implementation of technology in automated vehicles. Examples include rear cross-traffic warning (RCTW). A direct advantage of these systems is their automatic ability to work – allowing PwDs to utilize them regardless of ability, whilst a direct barrier would be how the sensors have been trained. These systems are typically trained with data where pedestrians are walking or stationary, are an average height, etc. – they do not account for those of short stature, those who use mobility aids, or those who may be a chair user \cite{molloy_addressing_2023}. Voice-controlled GPS guidance, driver-less systems, and simplified touchscreens are also included in this category. 

On the other hand, indirect advantages/disadvantages of automated vehicles are those that impact the life of a PwD outside the vehicle. These are typically societal and result from increased/decreased availability and use of accessible transportation. They are not directly related to the vehicle's internal system or functionalities. For example, access to independent transportation can drastically increase the social mobility of a PWD, allowing them to participate in activities with friends, in education, or on the job. This social inclusion would be considered an indirect, or societal, advantage. Alternatively, an indirect barrier may be the cost of automated vehicles. The cost is not associated with the use of these systems, but it does present a barrier as PwDs tend to have less income than the general population \cite{bureau_most_nodate}. These economic barriers are compounded by societal expectations and market forces.

Recognizing these categories provides critical context for regulatory decisions and highlights gaps in current accessibility efforts. By addressing both direct and indirect barriers while fostering targeted advantages, we can create a more inclusive and effective AV ecosystem.

\section{Current Regulations \& Policies}
\label{sec:regs}

\subsection{Global Frameworks}
Persons with Disabilities face barriers that limit their ability to live independently and fully participate in society. Accessibility, including access to transportation, is one of the critical components of the full and equal enjoyment of all human rights and fundamental freedoms.\cite{web:CRPD} There are two main international frameworks related to the equity of PwD: The Convention on the Rights of Persons with Disabilities (CRPD) and the Resolution WHA74.8 of the World Health Assembly\cite{web:WHA748}. Both primarily focus on indirect barriers.

The CRPD is a landmark treaty promoting the full and equal enjoyment of human rights by PwDs adopted by the United Nations General Assembly in 2006. It is a comprehensive human rights convention and an international development tool that aims to promote, protect, and ensure the full and equal enjoyment of all human rights and fundamental freedoms by Persons with Disabilities. It highlights the right to accessible transportation, with three key articles summarized in Table~\ref{table:CRPD_Articles}.

\begin{table}
  \centering
  \caption{CRPD Articles Related to Transportation}
  \label{table:CRPD_Articles}
  \begin{tabular}{c p{10cm}} 
    \toprule
    Article & Summary \\
    \midrule
    Article 9 & Emphasizes the importance of accessibility for Persons with Disabilities in various areas, including transportation. States Parties must take appropriate measures to ensure accessibility to transportation systems, facilities, and services, including identifying and eliminating barriers and obstacles. \\
    Article 19 & Emphasizes the right of Persons with Disabilities to live independently and be included in the community. Accessible transportation is crucial for enabling individuals with disabilities to access employment, education, healthcare, and other essential services, promoting their full inclusion and participation in society. \\
    Article 20 & Recognizes the right of Persons with Disabilities to personal mobility and emphasizes the importance of facilitating access to mobility aids, devices, assistive technologies, and forms of live assistance to enable their independence and full participation in society. \\
    \bottomrule
  \end{tabular}
\end{table}

World Health Assembly Resolution WHA74.8, adopted in May 2021, addresses the rights of Persons with Disabilities, including their access to transportation. While the resolution primarily promotes the health and well-being of Persons with Disabilities, it acknowledges the importance of transportation in facilitating their full participation in society. While WHA74.8 may not contain explicit statements regarding transportation rights for people with disabilities, it aligns with broader international efforts to promote accessibility, inclusion, and nondiscrimination for Persons with Disabilities across various domains, including transportation. Accessible transportation is essential to allow individuals with disabilities to exercise their rights to education, employment, healthcare, and community participation.

\subsection{USA Frameworks}
In the United States of America (USA), PwD transportation regulations include federal laws, state policies, and agency guidelines. While most address indirect barriers, some provide ambiguous guidance on direct barriers.

The National Highway Traffic Safety Administration (NHTSA) establishes automotive safety standards and AV policies but does not mandate accessibility-specific features. For instance, 49 CFR Part 595 Subpart C\cite{web:49CFRpart595} outlines vehicle adaptation guidelines for PwDs but does not address emerging AV technologies.

The Americans with Disabilities Act (ADA) \cite{web:ADAguide} prohibits discrimination against PwDs and requires public and private ground transportation to be accessible. While the ADA does not explicitly reference AVs, its principles could apply to AV-based public transit.

Since 2011, 22 states in the USA have passed legislation related to AVs based on the guidelines for Automated Driving Systems (ADS) \cite{web:ADSLeg} released by the NHTSA. While these laws sometimes address PwD accommodations, inconsistent state regulations risk fragmenting AV design, complicating accessibility and interstate travel.

Despite the presence of international frameworks like the Convention on the Rights of Persons with Disabilities (CRPD) and resolutions such as the World Health Assembly Resolution WHA74.8, and local laws such as the Americans with Disabilities Act (ADA) in the United States that mandate certain standards for transportation accessibility, enforcement remains inconsistent due to inadequate funding, limited awareness of accessibility standards, and weak accountability measures. As a result, even where strong legal frameworks exist, practical implementation often falls short, leaving gaps in transportation accessibility for PwDs.

\section{Proposed Collaborative Scoring System \& Rubric}
\label{sec:rub}
The lack of regulations addressing the accessibility of AVs for PwDs, the imbalance between indirect and direct barriers, and the prevailing ``benefit-afterthought'' narrative necessitate a structured evaluation system. To address this, we propose a scoring rubric that enables manufacturers, policymakers, and the public to assess the true accessibility of AVs. To the best of our knowledge, this is the first attempt of its kind.

The rubric fosters engagement between PwDs and AV developers by requiring survey feedback from PwDs, categorized based on functional ability rather than medical conditions. Participants assess their level of impairment before and after interacting with a new vehicle feature, ensuring that accessibility is measured in terms of real-world usability. As the advantages of AV systems should focus on improving quality of life and accessibility, it is important that the classification system maps to ability. Therefore, indicating who has a qualified disability for these surveys is a significant aspect of the scoring guide. Additionally, it is important to denote which features are important to measure in a vehicle, as too many could be overwhelming and too few could render the rubric useless. While designed for AVs, the system can also apply to nonautonomated vehicles.

\subsection{Disability Classification}
Disability classification varies across frameworks, including as defined by the International Classification of Functioning, Disability and Health (ICF), the Individuals with Disabilities Education Act (IDEA), and the International Paralympic Committee (IPC). IDEA covers 13 disability categories, with a focus on mental and learning disabilities \cite{noauthor_individuals_nodate}. However, the classification was created for children and learning and does not encompass a wide range of disabilities, ages, or levels of affect. The IPC covers athletes who participate in the Paralympics, with a focus on physical and mobility-related disabilities \cite{noauthor_ipc_nodate}. The IPC classification process and rules are intense and strict, with very little flexibility. Both IDEA and IPC classification systems would leave large groups of PwDs out of conversation, further limiting an already marginalized group. For the purposes of this paper, we suggest classification by use of the International Classification of Functioning, Disability and Health. 

The ICF was created by the World Health Organization (WHO) to measure disability in both individuals and populations \cite{noauthor_international_nodate}. It consists of four categories: body functions, activities and participation, environmental factors, and body structures. Each category has various subcategories with multiple classification codes like those used in the International Classification of Diseases (ICD). Excluding the environmental factors category, each classification code contains varying levels of extent or magnitude of the related impairment. 

For this rubric, we focus on the ICF's category ``Activities and Participation.'' 'Body functions' is excluded because the breakdown of each category is too broad to use appropriately. For example, 'Sensory functions and pain' breaks down into 'Seeing and related functions' and 'Pain', both broad in their descriptions and not indicative of any level of ability. 'Body Structures' is excluded for similar reasons: body structures present specific impairments of specific body structures, such as the 'Structure of eyeball' breaking into several structures, including the 'Cornea' and 'Iris'. Once again, the level of impairment is not related to the ability. Similarly, 'Environmental factors' is excluded.

'Activities and Participation' is defined at both the individual and societal level (making it easier to relate to both technical and social barriers / advantages), has levels of impairment, and defines a person's ability regardless of their body functions/structures/or medical association. However, this category defines many activities that automated vehicles – and vehicles in general – cannot give an indirect or direct advantage to. Furthermore, impairments relating to some categories may be improved by the indirect advantages of having an accessible vehicle (by which one may drive or be driven), collapsing to the Mobility activity category. This means that features in the vehicle are less likely to provide direct advantages or disadvantages to associated impairments of that activity or lack thereof and should not be included when discussing the accessibility of features. We limit our scoring rubric to three activities, each with 9 levels of impairment. Table~\ref{table:Activity_Categories} shows each activity category, whether it was included, and why. 

\begin{table}
  \centering
  \caption{Activity Categories Related to Vehicle Use}
  \label{table:Activity_Categories}
  \begin{tabular}{p{4cm} l p{8cm}} 
    \toprule
    Activity Category & Included (Yes/No) & Why? \\
    \midrule
    Learning and Applying Knowledge & No & Vehicles are not primarily utilized for learning or teaching. \\
    General Tasks and Demands & Yes & This activity includes carrying out daily routines, which vehicles may assist with when these routines involve leaving the home. Additionally, this category may include activities such as safely crossing the street in front of autonomous vehicles while utilizing a wheelchair, and other known safety risks that impede general tasks and demands. \\
    Communication & Yes & Modern vehicles commonly include interactive communicative features that connect the occupants and system components, or cellular devices. \\
    Mobility & Yes & This activity includes moving around using transportation, which is the primary usage of a vehicle. \\
    Self-care & No & Collapsed to Mobility. \\
    Domestic Life & No & This activity includes the acquisition of necessities, but this can be included in the General Tasks \& Demands category. All other activities within this category are not primary responsibilities of vehicles or vehicle travel. \\
    Interpersonal Interactions and Relationships & No & Features of a vehicle may make certain aspects of interpersonal relationships easier to accomplish, but it is not the primary goal of a vehicle’s feature to help create and maintain a relationship for any occupant. \\
    Major Life Areas & No & Collapsed to Mobility. \\
    Community, Social, and Civic Life & No & Collapsed to Mobility. \\
    \bottomrule
  \end{tabular}
\end{table}

There are 7 levels of impairment (labeled as 'Qualifiers' in the ICF) for each level of activity. Impairment levels are broken down into 'Performance' and 'Capacity' \cite{ustun_whos_2003, noauthor_icf_nodate}. 'Performance' denotes the 'lived experience' of a person, or how a person functions in day-to-day life with environmental factors, assistive devices, etc. 'Capacity' is what a person can do in a 'clinical' setting, or the maximum that they can do without environmental factors, outside help, or assistive devices. We are not interested in 'Capacity' measurements, as automated vehicles should be helpful in daily life. By removing assistive devices and environmental factors, we are not gaining a real-life understanding of how features may provide advantages or barriers to PwDs. This is especially impactful for those who utilize adapted vehicles and may be negatively affected by a new feature of a vehicle. Using the 'Performance' metric, adapted vehicle components and other assistive devices can still be used when testing the features of a vehicle; therefore, possible negative effects of the feature interacting with these adapted components may be discovered during the testing and survey process. 

The performance levels are indicated with a 'qpx' where x is the number of the level. x may be 0-4, representing NO, MILD, MODERATE, SEVERE, and COMPLETE performance difficulties, respectively. As qp0 would represent a person without an impairment, it is disregarded. The other two levels are qp8 and qp9, representing 'not specified' and 'not applicable' performance difficulties, respectively. These levels are also ignored for being ambiguous. Only impairment levels qp1, qp2, qp3, and qp4 are considered. To simplify the representative groups for scoring, PwDs belonging to the impairment groups qp1 \& qp2 are combined into a ``mild-moderate'' impairment group, while PwDs in qp3 \& qp4 are combined into a ``severe'' impairment group. These impairment groups and the rules associated with belonging to them are discussed in more detail in \nameref{sec:surveyscore}.

\subsection{Vehicle Feature Classification Framework}
\label{sec:vfeats}

Vehicle manufacturers categorize features differently, often for marketing purposes. Despite variations, features generally fall under three broad categories: Exterior, Interior, and Safety. However, these classifications lack consistency, leading to potential “feature manipulation,” where features are intentionally miscategorized to obtain better scores when assessing accessibility with the scoring system presented in \nameref{sec:surveyscore}. 

To address this, we propose a standardized feature classification framework, categorizing features based on Usability, Inclusivity, and Accessibility, inspired by human-centered design principles \cite{HumanDesign}. These categories ensure consistency and prevent misrepresentation in accessibility evaluations. The categories and their related feature descriptions are provided in Table ~\ref{table:Feature_Categorization}.

\begin{table}
  \caption{Feature Classification Proposal}
  \label{table:Feature_Categorization}
  \begin{tabular}{c p{10cm}} 
    \toprule
    Category & Description \\
    \midrule
    Usability & Features that enhance the driving experience by focusing on user-friendly products and meaningful solutions. These features were mainly designed to improve the system's efficiency, effectiveness, and satisfaction. Examples may include screen readers, voice commands, and adaptable interfaces. \\
    Inclusivity & Features that accommodate a wide range of users. These features were designed to consider human diversity by encompassing the needs of different user groups. Examples may include swivel seats, child safety technology, and mirror adjustment. \\
    Accessibility & Features designed to ensure users with different needs and abilities, including PwD, benefit from and can use them. Examples may include interfaces with touch, voice, and gesture inputs, sensory accommodations, and ramps. \\
    \bottomrule
  \end{tabular}
\end{table}

\subsection{Survey \& Scoring}
\label{sec:surveyscore}
The Vehicle Accessibility Rating System (VARS) assigns accessibility scores based on surveys completed by PwDs for each vehicle feature within the three classification categories outlined in \nameref{sec:vfeats}. To the authors’ knowledge, this is the first rubric of its kind.

PwDs participating in VARS surveys self-report into one of two impairment groups (Mild-Moderate or Severe). Participants can belong to multiple activity categories but must report only one impairment level per activity. For example, a participant may classify as Severe for Mobility, Mild-Moderate for Communication, and report no impairment for Domestic Life or General Tasks \& Demands—meaning they would only receive surveys for the first two activities. Participants can adjust their impairment classifications over time to reflect changes in ability. To ensure meaningful results, at least 20 survey responses are required per feature; otherwise, a Low Participation flag is applied to the score.

Each feature is evaluated based on its impact on PwDs’ ability to perform relevant activities. Participants are asked a single question per feature: \textit{``Does this feature change your level of impairment for this activity when using or interacting with the vehicle (as a driver, passenger, or pedestrian)?''} Responses are recorded on a scale from -5 to 5, where -5 indicates a significant negative impact, 0 indicates no change, and 5 indicates a strong positive impact. If a participant’s impairment level changes based on context (e.g., as a driver vs. passenger), they report the lowest impact score to ensure that potential accessibility barriers are not overlooked.

This scoring methodology is based on the concept of the Net Promoter Score, where high scores (9-10) are Promoters, medium scores (7-8) are Passives, and low scores (0-6) are Detractors \cite{noauthor_measuring_nodate}. The overall score is calculated by subtracting the percentage of Detractors from the percentage of Promoters (ignoring the Passives). To normalize our scoring rubric, and to show both positive and negative effects, VARS responses are instead categorized as:
\begin{itemize}
    \item \textbf{Promoters (4–5):} Feature significantly improves accessibility.
    \item \textbf{Passives (2–3):} Feature provides some benefit but is not transformative.
    \item \textbf{Detractors (-5 to 1):} Feature has little benefit or actively creates barriers.
\end{itemize}

We utilize this scale because minimally improving features may not be fully beneficial and are less likely to be beneficial for large groups of PwDs due to costs (economic, learning curve, etc.) involved in the implementation of the feature. Additionally, this scale allows us to balance those with higher levels of impairments and those with lower levels of impairment to prevent vehicle designers from creating and favoring accessible features that only work for those with minimal impairment. 


The 'no change' point on the -5 to 5 is dependent on the participants' current level of impairment. Participants who experience lower levels of impairment are likely to derive a greater benefit from a variety of features, whereas those with higher levels of impairment may require features tailored to their specific needs \cite{arias_differences_2020, thompson_expanding_2018, garcia_communication_2020}. In the same way, it should be easier to make a feature that positively affects a person with lower levels of impairment than a person with higher levels of impairment. For this reason, participants are asked to view their current level of impairment at different starting points or 'no change' sections of the scale. For those in the mild-moderate impairment group, their 'no change' level is -1, while the severe impairment group has a 'no change' level of 1. These ``no change'' starting points allow for features that positively impact those with high levels of impairment to have a greater impact on the score. When scoring, the starting levels are regarded as the same because they are both Detractors, but it is easier to move to a Promoter standpoint if a feature is accessible and advantageous for a person in the severe impairment group. This is to further encourage vehicle manufacturers to make accessible and advantageous features without favoring the lower impairment group simply due to their naturally wider range of accessibility. This balance also incentivizes manufacturers to adopt the rubric, as they will not be punished for imbalances or severe-impairment participant involvement. 

To score, the Detractors and Promoter percentages are totaled for each category of features, and for each activity group within that feature. The impairment groups for each activity and feature are combined on the VARS; the splitting of impairment groups is only to balance the scores as described previously. The percentage of Detractors (compared to the total number of participants for the activity) is subtracted from the percentage of Promoters and divided by 10, with the final vehicle score falling between -10 and 10. When the percentages are subtracted, the resulting number is rounded to the nearest whole number. For each feature category, this score can be averaged. However, to account for the differing numbers of features in each category, the vehicle's overall accessibility rating is the average of all feature scores; it is NOT the average of the three feature classification scores, as this would set feature classification groupings equal when they may drastically differ in size. This is also known as a 'weighted average', where the 'weight' for each feature score is the number of features that are included in it. If a ``Low Participation'' mark is noted on any of the activities, that mark is applied to the entire score.

A VARS score \textbf{above 0} suggests that a feature or vehicle design improves accessibility, making it beneficial for PwDs. A \textbf{negative score} indicates the feature or vehicle introduces barriers, suggesting modifications are needed. 

Table ~\ref{table:Feature_Scores} provides an example of a completed VARS evaluation for a vehicle with six features, yielding an overall accessibility score of \textbf{2.0}, indicating a generally beneficial design. If the score were calculated as an average of feature category scores, it would be \textbf{0.9}, demonstrating why summing individual feature scores provides a more accurate accessibility representation.

\begin{table}
  \centering
  \caption{Accessibility, Usability, and Inclusivity Feature Scores}
  \label{table:Feature_Scores}
  \begin{tabular}{p{4cm} p{4cm} lll} 
    \toprule
    \textbf{Feature} & \textbf{Activity (Total Participants)} & \textbf{Promoters} & \textbf{Detractors} &\textbf{Score/10} \\
    \midrule
    Accessibility Sample Feature 1 & General Tasks & 12 & 7 & 1.6 \\
                     & Communication & 5 & 0 & 2.5 \\
                     & Mobility & 25 & 2 & 5.0 \\
    Sample Feature 1 Score & \textit{Average} & & & 3.0 \\
 
    Accessibility Sample Feature 2 & General Tasks & 1 & 15 & -4.4 \\
                     & Communication & 4 & 4 & 0.0 \\
                     & Mobility & 1 & 31 & -6.5 \\
    Sample Feature 2 Score & \textit{Average} & & & -3.6 \\

    Accessibility Sample Feature 3 & General Tasks & 24 & 1 & 7.2 \\
                     & Communication & 0 & 0 & 0.0 \\
                     & Mobility & 37 & 0 & 8.0 \\
    Sample Feature 3 Score & \textit{Average} & & & 5.1 \\
    \midrule
    Accessibility Feature Score & \textit{Section Average} & & & 1.5 \\
    \midrule
    Usability Sample Feature 4 & General Tasks & 4 & 17 & -4.1 \\
                     & Communication & 6 & 8 & -1.0 \\
                     & Mobility & 0 & 41 & -8.9 \\
    Sample Feature 4 Score & \textit{Average} & & & -4.7 \\
    \midrule
    Usability Feature Score & \textit{Section Average} & & & -4.7 \\
    \midrule
   
    Inclusivity Sample Feature 5 & General Tasks & 16 & 0 & 5.0 \\
                     & Communication & 10 & 1 & 4.5 \\
                     & Mobility & 42 & 0 & 9.1 \\
    Sample Feature 5 Score & \textit{Average} & & & 6.2 \\
    Inclusivity Sample Feature 6 & General Tasks & 19 & 6 & 4.1 \\
                     & Communication & 18 & 0 & 9.0 \\
                     & Mobility & 21 & 3 & 3.9 \\
    Sample Feature 6 Score & \textit{Average} & & & 5.7 \\
    \midrule
    Inclusivity Feature Score & \textit{Section Average} & & & 6.0 \\
    \midrule
    \textbf{Vehicle Accessibility Score} & \textit{Weighted Section Average} & & & \textbf{2.0} \\
    \bottomrule
  \end{tabular}
\end{table}

Table. \ref{table:Activity_Scores} presents the \textit{Activity-Feature Scoring Summary}, reorganizing data by activity category. This format is particularly useful for PwDs, allowing them to assess how features impact their specific impairments. For example:
\begin{itemize}
    \item \textbf{Features 2 and 4} introduce barriers, particularly for individuals with *General Task \& Mobility Impairments*.
    \item \textbf{Feature 3} improves accessibility for individuals with *General Task \& Mobility Impairments* but offers little benefit for *Communication Disabilities*.
    \item Removing barriers (e.g., Features 2 \& 4) would improve the overall vehicle score.
\end{itemize}

\begin{table}
  \centering
  \caption{Activity-Based Feature Summary}
  \label{table:Activity_Scores}
  \begin{tabular}{l l l l} 
    \toprule
         \textbf{Activity (Total Participants)} & & & \\
      \textbf{Features} &
    \textbf{Promoters} & \textbf{Detractors} & \textbf{Score/10} \\
    \midrule 
     General Tasks \& Demands (32) Activity &&&\\
    \midrule 
    Sample Feature 1 & 12 & 7 & 1.6 \\
    Sample Feature 2 & 1 & 15 & -4.4 \\
    Sample Feature 3 & 24 & 1 & 7.2 \\
    Sample Feature 4 & 4 & 17 & -4.1 \\
    Sample Feature 5 & 16 & 0 & 5.0 \\
    Sample Feature 6 & 19 & 6 & 4.1 \\
    \midrule
    \textbf{General Tasks \& Demands Activity Score} & & & \textbf{1.6}\\
    \midrule
    Communication (20) Activity &&&\\
    \midrule 
    Sample Feature 1 & 5 & 0 & 2.5 \\
    Sample Feature 2 & 4 & 4 & 0.0 \\
    Sample Feature 3 & 0 & 0 & 0.0 \\
    Sample Feature 4 & 6 & 8 & -1.0 \\
    Sample Feature 5 & 10 & 1 & 4.5 \\
    Sample Feature 6 & 18 & 0 & 9.0 \\
    \midrule
    \textbf{Communication Activity Score} & & & \textbf{2.5}\\
    \midrule
     Mobility (46) Activity  &&&\\
    \midrule 
    Sample Feature 1 & 25 & 2 & 5.0 \\
    Sample Feature 2 & 1 & 31 & -6.5 \\
    Sample Feature 3 & 37 & 0 & 8.0 \\
    Sample Feature 4 & 0 & 41 & -8.9 \\
    Sample Feature 5 & 42 & 0 & 9.1 \\
    Sample Feature 6 & 21 & 3 & 3.9 \\
    \midrule
    \textbf{Mobility Activity Score} & & & \textbf{1.8}\\

    \bottomrule
  \end{tabular}
\end{table}

The VARS system balances the needs of PwDs and manufacturers by ensuring accessibility is assessed both systematically and equitably. The \textit{Vehicle Accessibility Rating System} should be used in vehicle evaluations, while the \textit{Activity-Feature Scoring Summary} provides PwDs with actionable insights. Displaying both formats together allows all stakeholders—consumers, manufacturers, and policymakers—to make informed decisions about vehicle accessibility.

\section{Conclusion}
VARS is a first-of-its-kind accessibility scoring system for automated vehicles with enough flexibility to be retroactively applied to preexisting, nonautomated, commercial vehicles. The novelty of the proposed solution limits its practical usage without user-testing; however, it is the first attempt at creating a universal, accessible scoring rubric that encourages communication and collaboration with PwDs during the design phase of vehicle manufacturing. Future efforts will focus on empirically evaluating the rubric, in collaboration with Persons with Disabilities (PwDs), to iteratively refine the scoring guide until it reaches a standard suitable for regulatory adoption.

The integration of AV systems has transformative potential to improve mobility, particularly for people with disabilities. The development of accessible designs represents a critical intersection of ethical, legal, and social imperatives to support independence, safety, and equality. It also reinforces the principles of inclusivity and nondiscrimination to ensure that the benefits of new technologies are universally shared.

By integrating VARS into vehicle design and evaluation processes, manufacturers, policymakers, and researchers can work toward an inclusive transportation future. As AV adoption grows, ongoing collaboration between industry stakeholders and PwDs will be essential to refining accessibility measures, fostering innovation, and ensuring that the benefits of automated transportation are widely and equitably shared.

\begin{acks}
This work began as a student project and received no financial support.
\end{acks}

\bibliographystyle{ACM-Reference-Format}
\bibliography{biblio}

\end{document}